\documentclass[12pt,thmsa,epsf]{article}
\usepackage{amssymb}
\usepackage{epsf}

\input{tcilatex}
\begin{document}

\author{R. L. Ingraham \\
%EndAName
\textit{Department of Physics, New Mexico State University }\\
\textit{Box 30001, Dept. 3D, }\\
\textit{Las Cruces, NM\ 88003-8001, U.S.A.}}
\title{\textbf{Particle masses and the fifth dimension, final version }}
\date{}
\maketitle

First we argue in an informal, qualitative way that it is natural to enlarge
space-time to five dimensions to be able to solve the problem of elementary
particle masses. Several criteria are developed for the success of this
program. Extending the Poincar\'{e} group to the group \textbf{C} of all
angle-preserving transformations of space-time is one such scheme which
satisfies these criteria. Then we show that the field equation for spin 1/2
fermions coupled to a self-force gauge field predicts mass spectra of the
desired type: for a certain range of a key parameter (Casimir invariant) a
three-point mass spectrum which fits the ``down'' quarks $d,s,$ and $b$ to
within their experimental bounds is obtained. Reasonable values of the
coupling constant (of QCD magnitude) and the range of the spatial wave
function (a few fermis) also result. Compatibility with the electroweak
theory is also discussed.

\bigskip

\noindent \textbf{1. INTRODUCTION}

\medskip

A theory of elementary particle masses which predicts the masses that we see
in nature is lacking in present day particle physics. The Standard Model
appeals to the Higgs mechanism. But even granting that the Higgs particle
exists, successful fits must wait on the measurement of various unknown
parameters [1]. String theories claim to be able to predict these masses in
principle, but they are still far from delivering quantitative numbers at
their present stage [2,3].

First, some informal, qualitative remarks may be helpful to motivate the
main idea of this paper. The idea that predicting particle masses should
involve enlarging 4-D (``four-dimensional'') space-time (coordinates $%
x^{^\mu }=\{x,y,z,x^o\equiv ct\}$ by a single new dimension, call it $%
\lambda $, seems very natural. The equal status of momentum $\mathbf{p}$,
energy $E$, and mass $m$ in the free particle relation 
\begin{equation}
\mathbf{p}^2-E^2+m^2=0  \tag{1}
\end{equation}
suggests that in 5-D position space $\lambda $ should be conjugate to $m$,
just as $\mathbf{r}$ is conjugate to $\mathbf{p}$ and $t$ is conjugate to $E$%
. (We shall use units $c=1$ in this paper.) And further, that the field
equation for the field $\phi (x^{^\mu },\lambda )$ of a free scalar boson,
say, should be something like 
\begin{equation}
(\bigtriangledown ^2-\partial ^2/\partial t^2+\partial ^2/\partial \lambda
^2)\phi (x^{^\mu },\lambda )=0\text{ ,}  \tag{2}
\end{equation}
with the solution 
\begin{equation}
\phi (x^{^\mu },\lambda )=const\times \exp [i(\mathbf{p}\cdot \mathbf{r}%
-Et\pm m\lambda )]  \tag{3}
\end{equation}
with the constraint (1) on the constants $\mathbf{p}$, $E$, and $m$.

However, this first try is too naive for several reasons. First, the new
dimension $\lambda $ is simply grafted onto space-time, uncritically
assuming that the enlarged space is still flat (\textit{cf}. Eq. (2)). The
symmetry group of Eq. (2) and of the corresponding 5-D metric 
\begin{equation}
dS^2=d\mathbf{r}^2-dt^2+d\lambda ^2  \tag{4}
\end{equation}
is the set of 5-D rotations and translations. But this group preserves
nothing significant in space-time. One would like the new symmetry group to
be related to some structure defined in space-time alone, to preserve some
geometric entity of space-time.

The second reason that Eq. (2) is too naive is that the mass spectrum is
continuous: $0<m<\infty $. But the whole mystery of particle mass spectra is
that they consist of a few points with non-uniform spacing! Clearly a 
\textit{perfectly} free particle field equation like (2) can never predict
mass spectra of this type. We suggest that there should always be a
self-force acting on the particle, whether or not it is acted on by external
forces. The self-force must certainly involve the new coordinate $\lambda $,
conjugate to mass.

The third reason that Eq. (2) is too naive is that it was simply written
down \textit{ad hoc} without any regard for the symmetry group of the new
5-D space. But as Bargmann and Wigner showed many years ago [4], the
particle field equations now accepted --- the scalar boson equation, the
Dirac equation for spin 1/2 fermions, the photon field equation, etc. ---
correspond to the irreducible unitary representations of the Poincar\'{e}
group $\mathbf{P}$, labelled by its two Casimir invariants spin $j$ and mass 
$m$, which uniquely fix these equations.\ Therefore the new symmetry group
should have been chosen first, in accordance with the first criterion above,
and then the field equations of the various particle species determined by
its IUR's.

Back to the first criterion: the present kinematical symmetry group of
space-time is the Poincar\'{e} group $\mathbf{P}$, which preserves the
space-time length element $ds^2=g_{\mu \nu }dx^{^\mu }dx^{^\nu }\equiv d%
\mathbf{r}^2-dt^2$. One is thus motivated to search for the simplest and
smallest extension of $\mathbf{P}$ which preserves something geometrical in
space-time and has $\mathbf{P}$ as a subgroup. An immediate candidate is the
group $\mathbf{C}$ which preserves space-time angle. By Liouville's Theorem
[5] $\mathbf{C}$ is a 15-parameter Lie group composed of the 10-parameter
subgroup $\mathbf{P}$, which preserves space-time length (and therefore
space-time angle) augmented by a 5-parameter set of transformations which
preserve space-time angle but not length.

To answer an expected immediate objection: of course $\mathbf{C}$'s
transformations cannot act just on the 4-D space-time, with its length
metric $ds^2=g_{\mu \nu }$ $dx^{^\mu }dx^{^\nu }$, because angle-preserving
transformations of space-time do not in general preserve the length, and
thus $\mathbf{C}$ would not be the symmetry group of this metric. (This was
Einstein's reason for rejecting the group $\mathbf{C}$, see [6].) The way to
introduce the group $\mathbf{C}^n$ of \textit{conformal} $(\equiv $
angle-preserving) transformations of $n$-dimensional euclidean space $E^n$
of coordinates $x^{^\mu },\mu =1,2\cdots n$, was well-known to the great
geometers of the nineteenth century (F. Klein, Liouville, M\"{o}bius, Lie 
\textit{et al.}) some 150 years ago, but seems unknown today, at least to
modern theoretical physicists. In brief, one introduces the $(n+1)$%
--dimensional space of spheres in $E^n$ characterized by their centers $%
x^{^\mu }$ and radii $x^{n+1}$. The group $\mathbf{C}^n$ is then that group
of transformations $x^{\prime ^\alpha }=f^{^\alpha }(x^1,x^2,\cdots
x^n,x^{n+1}),$ $\alpha =1,2,\cdots n,n+1,$ which preserve the angle $\theta $
under which two spheres $x^{^\alpha }$ and $y^{^\alpha }$ intersect, see
Fig. 1. For infinitesimally close spheres $y^{^\alpha }=x^{^\alpha
}+dx^{^\alpha }$ one gets [7] 
\begin{equation}
d\theta ^2=(x^{n+1})^{-2}[(dx^1)^2+(dx^2)^2+\cdots (dx^n)^2-(dx^{n+1})^2]%
\text{ .}  \tag{5}
\end{equation}
(This is nothing but the Law of Cosines, familiar from plane geometry class
in high school.) The expression (5) defines the metric (dimensionless angle
metric) of the appropriate $(n+1)$-dimensional Riemannian space which has
the conformal group $\mathbf{C}^n$ as its symmetry group. It turns out that
this space is not flat but is of constant curvature. All of this is
explained in exhaustive detail elsewhere [7].

Thus for the pseudo-euclidean space-time with $n=4$ we get the 5-dimensional
space with metric 
\begin{equation}
d\theta ^2=-\lambda ^{-2}(d\mathbf{r}^2-dt^2+\sigma d\lambda ^2),\sigma =\pm
1\text{ ,}  \tag{6}
\end{equation}
with $\{x^1,x^2,x^3,x^4\}$ and $x^5$ renamed $\{x^1,x^2,x^3,x^0\}$ and $%
\lambda $ respectively. Of course the ``sphere'' $x^{^\alpha }$ is the
hyperboloid $g_{\mu \nu }(\xi ^{^\mu }-x^{^\mu })(\xi ^{^\nu }-x^{^\nu
})+\sigma \lambda ^2=0$ as a real locus. The sign $\sigma $, that is,
whether the fifth dimension is spacelike $(\sigma =+)$ or timelike $(\sigma
=-)$ is left open for the moment.

This concludes the informal, qualitative part of this Introduction.

We show here how the field equation for spin 1/2 fermions in five dimensions
coupled to a self-force dependent on the fifth coordinate predicts point
mass spectra of just a few points and non-uniform spacing. If the Casimir
invariant of this particular irreducible unitary representation has a
certain range, it is a 3-point spectrum for isospin up or down. The spectrum
is consistent with the experimental bounds on the isospin-down quarks $d$, $%
s $, and $b$ for values of the coupling constant $\alpha $ of order unity
and range $\kappa ^{-1}$ of the spatial wave functions of a few fermis.

To avoid a possible confusion at the outset: this 5-D theory has nothing to
do with the Kaluza or Kaluza-Klein theories. The enlargement of space-time
to a five-dimensional manifold is forced, not arbitrary, if the conformal
group is demanded as the basic kinematical symmetry group [7]. This fifth
coordinate $\lambda $ is conjugate to mass just as position and time are
conjugate to momentum and energy. \ Partial derivatives with respect to $%
\lambda $ replace mass terms in fermion and boson field equations. In
solutions of gauge boson field equations $\lambda $ plays the role of a
microscopic length ``parameter'' which modifies the usual space-time
causality of point particles. It gives point particles a structure or
extension in a certain sense [7].

We argue in this paper that this five-dimensional extension of special
relativity (``conformal relativity'') is the natural framework for a theory
of elementary particle mass.\ The results obtained here are promising but
are only a first step; the main problem is the exact form of the
quantum-mechanical self-force. Some extra points, including a puzzle, are
made in the concluding remarks. These also include an argument that the 5-D
theory gives a theoretical basis for some features of the electroweak theory
which were postulated on the basis of experiment alone.

\bigskip

\noindent \textbf{2. SOME\ BACKGROUND}

\medskip

As explained in the Introduction, the metric of conformal relativity is [7] 
\[
d\theta ^2=-\lambda ^{-2}(dx^2+\sigma \text{ }d\lambda ^2)\text{ ,} 
\]
\begin{equation}
dx^2\equiv g_{\mu \nu }dx^{^\mu }dx^{\nu \text{ }},\qquad \mu ,\nu =0,1,2,3;%
\text{ }x^5\equiv \lambda ;\qquad \sigma =\pm \text{ ,}  \tag{2.1}
\end{equation}
where $d\theta $ is the infinitesimal angle under which spheres $(x^{^\mu
},\lambda )$ and $(x^{^\mu }+dx^{^\mu },$ $\lambda +d\lambda )$ intersect.
We use the metric $-g_{00}=g_{11}=g_{22}=g_{33}=+1$. Whether the extra
dimension is spacelike $(\sigma =+)$ or timelike $(\sigma =-)$ is not yet
clear, or maybe both occur. The ranges of the coordinates are $-\infty
<x^{^\mu }<+\infty $ as usual, and $0<\lambda <\infty $ (or possibly $0<\mid
\lambda \mid <\infty )$. The metric is singular if $\lambda =0$, so $\lambda
=0$ is excluded from physical space, which is of course consistent with the
action of the conformal group \textbf{C} [7]. We call these two 5-D
Riemannian spaces (2.1) K$_{+}$ and K$_{-}$ (after Felix Klein).

The field equation for spin 1/2 fermions in the \textbf{C}-covariant theory
is$\footnote{%
Eq. (2.2) here is Eq. (4.3) of the second article of Ref. [8], where $\nu
\equiv -(4/9)q_3$. Note that these articles considered only the case $\sigma
=+$. Much of the physical discussion there is dated.}$ [8] 
\begin{equation}
(\gamma ^{^\alpha }\nabla _{_\alpha }+\gamma \beta _7\nu )\text{ }\psi =0%
\text{ , }\qquad \nabla _{_\alpha }\equiv \stackrel{\gamma }{\nabla }%
_{_\alpha }-ig\text{ }A_{_\alpha }\text{ .}  \tag{2.2}
\end{equation}
Here the six anticommuting $\gamma $-matrices obey 
\begin{equation}
\gamma _{_\alpha }\gamma _{_\beta }+\gamma _{_\beta }\gamma _{_\alpha
}=2\gamma _{\alpha \beta }\text{ }\mathbf{1}\text{, }\gamma _{_\alpha
}\gamma +\gamma \gamma _{_\alpha }=0\text{, }\gamma ^2=\mathbf{1}\text{ ,} 
\tag{2.3a}
\end{equation}
\begin{equation}
\beta _7\equiv i\lambda ^5\gamma _1\gamma _2\gamma _3\gamma _0\gamma
_5\gamma \text{ ; }\alpha ,\beta =0,1,2,3,5\text{ ,}  \tag{2.3b}
\end{equation}
where $\gamma _{\alpha \beta }$ is the angle metric (2.1). Indices are
raised and lowered with this metric. $\stackrel{\gamma }{\nabla }_{_\alpha }$
is the covariant derivative on spinors $\psi $ which fixes the spin algebra $%
\gamma _{_\alpha },\gamma $. (Note that the spaces $K_{_\sigma }$ are not
flat, so that covariant derivatives occur in field equations.) We consider
here only a $U(1)$ internal symmetry with gauge boson $A_{_\alpha }$. The
equation (2.2) is uniquely fixed by requiring that the solutions $\psi $
span an irreducible unitary representation $(IUR)$ of \textbf{C}. The
parameter $\nu $ is a Casimir invariant for this $IUR$, and Eq. (2.2) is the
sole independent condition for spin 1/2 [8]. The six $\gamma _{_\alpha
},\gamma $ are $8\times 8$ and $\psi $ is an 8-spinor because eight is the
minimum dimension allowed for a matrix representation of the algebra (2.3a).
When the spin connection is inserted, Eq. (2.2) reduces to

\[
(\tilde{\gamma}\cdot D+\tilde{\gamma}^5D_5+2\sigma \lambda \tilde{\gamma}%
_5+\nu \beta _7)\psi =0\text{ ,} 
\]
\begin{equation}
\tilde{\gamma}_{_\alpha }\equiv \gamma \gamma _{_\alpha }\text{ , }\qquad
D_{_\alpha }\equiv \partial _{_\alpha }-igA_{_\alpha }\text{ ,}  \tag{2.4}
\end{equation}

\noindent where the $^{\bullet }$ will always mean the 4-D scalar product $%
\tilde{\gamma}\cdot D\equiv \tilde{\gamma}^{^\mu }D_{_\mu }$. Note that $%
D_{_\alpha }$ involves the ordinary partial derivative $\partial _{_\alpha }$
; the third term in Eq. (2.4) comes from the spin connection.

To be able to calculate with Eq. (2.4) a representation of the six $8\times
8 $ matrices $\gamma _{_\alpha },\gamma $ must of course be chosen. We
choose $\gamma _{_\alpha }=\gamma \stackrel{\sim }{\gamma _{_\alpha }}$ and 
\begin{equation}
\stackrel{\sim }{\gamma }_{_\mu }=\lambda ^{-1}\left( 
\begin{array}{ll}
-\gamma _{_\mu } & 0 \\ 
0 & \gamma _{_\mu }
\end{array}
\right) ,\stackrel{\sim }{\gamma _{\text{ }}}_5\text{ }=\lambda ^{-1}\left( 
\begin{array}{ll}
h & 0 \\ 
0 & -h
\end{array}
\right) ,\gamma =\left( 
\begin{array}{ll}
0 & 1 \\ 
1 & 0
\end{array}
\right) .  \tag{2.5a}
\end{equation}
The $\stackrel{\sim }{\gamma }^{\text{ }\alpha }$are obtained by raising the
indices with the metric (2.1). For the $4\times 4$ $\gamma _{_\mu },$ $h,$
and $1$ in these matrices, see Eq. (2.6). Then $\beta _7$, Eq. (2.3b), is 
\begin{equation}
\beta _7=\left( 
\begin{array}{ll}
1 & 0 \\ 
0 & -1
\end{array}
\right) \text{ .}  \tag{2.5b}
\end{equation}
It can be shown (unpublished) that by comparing a Lagrangian for the spin
1/2 field equation (2.4) with the Lagrangian for the electroweak theory
([1],\ Chap. 7) that we can identify the upper and lower 4-spinors in the
8-spinor $\psi $ as the $T_3=+1/2$ and $-1/2$ components of the isodoublets
of the electroweak theory in this representation. In fact, the whole
electroweak theory can be reproduced. More on this in \textbf{Sec. 4.}
Therefore we call the representation (2.5) the EW (electroweak)
representation. The field equation (2.4) written in the $EW$ representation
splits cleanly into wave equations for the $T_3=+1/2$ and $-1/2$ components
(there is no coupling between these fields) and further, these wave
equations are identical.

This common wave equation for the case $\sigma =-$ is 
\[
\{\gamma \cdot (\partial -ig\text{ }A)-ih(\partial _5-ig\text{ }A_5)+(\nu
+2ih)/\lambda \}\psi =0\text{ ,} 
\]
\begin{equation}
\gamma _{_\mu }\gamma _{_\nu }+\gamma _{_\nu }\gamma _{_\mu }=2g_{\mu \nu }%
\mathbf{1}\text{ , }h\equiv i\gamma _1\gamma _2\gamma _3\gamma _0\text{ .} 
\tag{2.6}
\end{equation}
Here the $\gamma _{_\mu }$ are the usual $4\times 4$ constant $\gamma $%
-matrices, $\psi $ is now a $4$-spinor, and $h$ is the handedness operator
(usually called $\gamma _5$ in the literature): $h\psi _L=-\psi _L$, $h\psi
_R=+\psi _R$ for left and right-handed spinors.

The field equation for the gauge boson $A_{_\alpha }$ is 
\begin{equation}
\stackrel{\gamma }{\nabla }_{_\alpha }F_{\;\;_\beta }^{^\alpha }=0\text{ , }%
\qquad F_{\alpha \beta }\equiv \partial _{_\alpha }A_{_\beta }-\partial
_{_\beta }A_{_\alpha \text{ }}\text{.}  \tag{2.7}
\end{equation}
These are reduced to a set of partial differential equations for the $5$%
-vector $A_{_\alpha }$ in Ref. [7].

\bigskip

\noindent \textbf{3. FERMION\ MASS\ SPECTRUM\ FOR\ A\ TIMELIKE\ FIFTH\
DIMENSION}

\medskip

We look at stationary states: $\psi (t,\mathbf{r},\lambda )=e^{-iEt}g(%
\mathbf{r},\lambda )$ of Eq. (2.6). If we insert a \textit{self-force} $%
A_{_\alpha }^{SF}$ and solve for a resting spin 1/2 fermion, the energy
spectrum should be the mass spectrum: $E=M$. The self-force should certainly
involve the fifth coordinate $\lambda $, so we adopt provisionally 
\begin{equation}
A_0^{SF}=-g^{\prime }/\lambda \text{, other }A_{_\alpha }^{SF}\equiv 0\text{
.}  \tag{3.1}
\end{equation}
More on this in \textbf{Sec. 4.} Then the equation becomes 
\begin{equation}
\left\{ \gamma ^0(M-\alpha /\lambda )+i\mathbf{\gamma }\cdot \mathbf{%
\partial }+h\partial _{_\lambda }+(i\nu -2h)/\lambda \right\} \text{ }g(%
\mathbf{r},\lambda )=0\text{. }  \tag{3.2}
\end{equation}
Here $\alpha \equiv g^{\prime }g$ ($g^{\prime }=g$ is natural for a \textit{%
self}-force, but we leave this open for generality.) Consider $s$-states $%
g(r,\lambda )$ only; then $i\mathbf{\gamma }\cdot \mathbf{\partial }$
becomes $i\gamma _r\partial _r$ where $\gamma _r\equiv \mathbf{\gamma }\cdot 
\mathbf{n}$, $\mathbf{n}$ a unit 3-vector. We seek a separable solution in $%
r $ and $\lambda $, so take $g(r,\lambda )=e^{-\kappa r}g(\lambda )$ with $%
\kappa $ real and positive. Eq. (3.2) then reduces to the ordinary
differential equation in $\lambda $%
\begin{equation}
\left\{ \gamma ^0(M-\alpha /\lambda )-i\kappa \gamma _r+h\partial _{_\lambda
}+(i\nu -2h)/\lambda \right\} g(\lambda )=0.  \tag{3.3}
\end{equation}
The solution is given in the Appendix. It is formally very similar to the
solution of the Dirac equation for the relativistic hydrogen atom [9] with $%
\lambda $ and the mass levels of the particle playing the roles of $r$ and
the hydrogenic energy levels, respectively. (The spectrum is very different
however.) The mass spectrum is 
\[
M_{(n^{\prime },\tau )}/\kappa =\mid S_{_\tau }+n^{\prime }\mid \diagup
[\alpha ^2-(S_{_\tau }+n^{\prime })^2]^{1/2} 
\]
\begin{equation}
n^{\prime }=0,1,2,3,\cdot \cdot \cdot \text{ }\qquad ,\text{ }\tau =\pm 
\text{ }\qquad ,  \tag{3.4a}
\end{equation}
\begin{equation}
S_{_\tau }\equiv \tau (\alpha ^2-\nu ^2)^{1/2}\text{ ,}  \tag{3.4b}
\end{equation}
\begin{equation}
S_{_\tau }+n^{\prime }\text{ has the sign of }\alpha \equiv g^{\prime }g%
\text{ ,}  \tag{3.4c}
\end{equation}
\pagebreak

\text{norm restriction}\footnote{
For the 4-spinor $\psi $ the norm is $\mid \mid \psi \mid \mid ^2\equiv \int 
d^3r\int_0^\infty d\lambda \lambda ^{-4}\stackrel{-}{\psi } 
 \gamma _0 \psi $, $t=const$., where $\gamma _0$ is the constant 
$4\times 4$ matrix. For this ``bound'' solution we require 
$|| \psi ||^2<\infty $. 
The bound (3.4d) on $\gamma $ comes from requiring the $\lambda$-integral 
to converge at its lower limit $\lambda =0$.}:
\vspace*{-.3in}
\begin{equation}(\alpha^2 -\nu ^2)^{1/2}<1/2\text{ for }\tau =-\text{ .}  
\tag{3.4d}\end{equation}

One can see first in a general sort of way that this is a finite point
spectrum: when the radicand in the denominator of Eq. (3.4a) goes negative,
the spectrum ends. In fact, if we choose $\gamma \equiv (\alpha ^2-\nu
^2)^{1/2}$ as a convenient independent variable (do not confuse this $\gamma 
$ with the matrix $\gamma $ in Eq. (2.3)!) and set $F(\gamma ;n^{\prime
},\tau )\equiv \alpha ^2-(S_{_\tau }+n^{\prime })^2$, we get, on expanding
and cancelling etc. 
\begin{equation}
F(\gamma ;n^{\prime },\tau )=-2n^{\prime }\tau \gamma +\nu ^2-n^{\prime 2}%
\text{ .}  \tag{3.5}
\end{equation}
Now choose $g^{\prime }=g$, or $\alpha \equiv g^2>0,$ as seems natural. Then
the necessary and sufficient conditions for a spectral point $(n^{\prime
},\tau )$ are 
\begin{equation}
\gamma <(\nu ^2-n^{\prime 2})\diagup 2n^{\prime }\text{ , }\tau =+\text{ ; }%
\gamma >(n^{\prime 2}-\nu ^2)\diagup 2n^{\prime },\text{ }\tau =-\text{ , } 
\tag{3.6a}
\end{equation}
\begin{equation}
\gamma <n^{\prime }\text{ for }\tau =-\text{ }\qquad ,  \tag{3.6b}
\end{equation}
\begin{equation}
\gamma <1/2\text{ for }\tau =-\qquad .  \tag{3.6c}
\end{equation}
These are respectively from $F(\gamma ;n^{\prime }\tau )>0$, Eq. (3.4c) for $%
\alpha >0$, and Eq. (3.4d).

In modern particle theory there are three families (isodoublets) of quarks
and three of leptons. Relevant to this, the following theorem can be proved
from the conditions (3.6a, b, c):

\medskip

\textit{Theorem.} There are three and only three mass levels if and only if $%
1<\nu ^2<2$. These levels are $(n^{\prime },\tau )=(0,+)$, $(1,-)$, and $%
(1,+)$.

The mass spectrum written in terms of $\gamma $ is 
\begin{equation}
M_{(n^{\prime },\tau )}\diagup \kappa =(\tau \gamma +n^{\prime })\diagup
(-2n^{\prime }\tau \gamma -n^{\prime 2}+\nu ^2)^{1/2}  \tag{3.7}
\end{equation}
from just above. Thus for the three levels $(0,+)$, $(1,-)$, and $(1,+)$ we
get 
\begin{equation}
\begin{array}{lll}
M_{(0,+)}\diagup \kappa & = & \gamma \diagup \mid \nu \mid \text{ ,} \\ 
&  &  \\ 
M_{(1,-)}\diagup \kappa & = & (1-\gamma )\diagup (2\gamma -1+\nu ^2)^{1/2}%
\text{ ,} \\ 
&  &  \\ 
M_{(1,+)}\diagup \kappa & = & (1+\gamma )\diagup (-2\gamma -1+\nu ^2)^{1/2}%
\text{ ,} \\ 
&  &  \\ 
1<\nu ^2<2 & , & 0<\gamma <1/2\text{ }\qquad .
\end{array}
\tag{3.8}
\end{equation}
Then from these expressions one can deduce that the \textit{only}
possibility that one mass is much greater than the other two is 
\begin{equation}
\nu ^2=1+2\gamma +\varepsilon \text{ , }\qquad 0<\varepsilon <<1\text{ ,} 
\tag{3.9}
\end{equation}
in which case $M_{(1,+)}$ is the large one. (This assumes $\varepsilon
<<\gamma .$)

\medskip

\textit{Fitting the quarks}. We try to fit the $T_3=-1/2$ set of quarks $d$, 
$s$, and $b$. The experimental mass limits in $MeV$ are [10] 
\begin{equation}
M_d=3-9\text{, }M_s=60-170\text{, }M_b=4100-4400.  \tag{3.10}
\end{equation}
So we adopt the value (3.9) for $\nu ^2$ and identify $(1,+)\equiv b$. Next,
inserting $\nu ^2$ (3.9) into the mass formulae (3.8) and neglecting $%
\varepsilon $ in $(0,+)$ and $(1,-)$, we get the ratio 
\begin{equation}
M_{(1,-)}\diagup M_{(0,+)}=(1-\gamma )(1+2\gamma )^{1/2}\diagup 2\gamma
^{3/2}\text{ .}  \tag{3.11}
\end{equation}
It can be checked that this ratio is always $>1$ for $0<\gamma <1/2$, so we
choose $(1,-)\equiv s$ and $(0,+)\equiv d$. Now equate the ratio (3.11) to $%
M_s/M_d$, using the average values $M_d=6$ $MeV$ and $M_s=115$ $MeV$. The
resulting equation 
\begin{equation}
(1-\gamma )(1+2\gamma )^{1/2}=38.4\text{ }\gamma ^{3/2}  \tag{3.12}
\end{equation}
has the solution $\gamma \approx .088$. Finally, to determine $\varepsilon $%
, set the theoretical and experimental ratios $M_b/M_d$ equal. This gives 
\begin{equation}
(1+\gamma )\mid \nu \mid \diagup \gamma \varepsilon ^{1/2}=(M_b/M_d)\text{ }%
_{\exp tl\text{ }}\text{.}  \tag{3.13}
\end{equation}
Insert $\gamma =.088$ and $\mid \nu \mid =1.088$ and use the minimum value $%
4100/9\approx 455$ for the ratio on the right to get the maximum size of $%
\varepsilon $. This gives $\varepsilon _{\max }\approx 8.7\times 10^{-4}$,
and verifies our assumption $\varepsilon <<\gamma $.

The values of the coupling constant $\alpha $ and the range $\kappa ^{-1}$
of the spatial wave functions are also of interest. We can evaluate $\kappa $
from $\kappa (M_{(n^{\prime },\tau )}/\kappa )=(M_q)_{\exp tl}$. If we use
the same average values for $M_d$ and $M_s$ as used above to determine $%
\gamma $, we will get the same $\kappa $ for either $(1,-)$ or $(0,+)$.
Choose $(0,+).$%
\[
\kappa \gamma /\mid \nu \mid \text{ }=.081\kappa =6\text{ }MeV\Rightarrow
\kappa =74.2\text{ }MeV\text{, } 
\]
which gives $\kappa ^{-1}\approx 200/74.2\approx 2.7$ $f$. Also $\alpha
^2=\gamma ^2+\nu ^2\approx 1.18$, or $\alpha \approx 1.09$, which suggests a
self-force of QCD origin.

In summary, a fit to the three isospin-down quarks $d$, $s$, and $b$ has
been obtained as the levels 
\begin{equation}
(0,+)\equiv d\text{, }\qquad (1,-)\equiv s\text{ ,}\qquad (1,+)\equiv b 
\tag{3.14a}
\end{equation}
for the Casimir invariant $\nu ^2\approx 1.176$ and the reasonable values of
the physical parameters 
\begin{equation}
\alpha \approx 1.09\text{ and }\kappa ^{-1}\approx 2.7\text{ }f\text{ .} 
\tag{3.14b}
\end{equation}
Of course nearby values of these parameters will also give a fit owing to
the wide latitude (3.10) in the experimental masses.

\bigskip

\noindent \textbf{4. CONCLUDING\ REMARKS}

\medskip

A further characteristic of this theory necessary in any theory of mass
should be mentioned. In inelastic scattering of elementary particles, energy
and momentum are conserved but mass is not. Thus in any theory which unifies
these quantities in some sense mass must be qualitatively different from
energy and momentum and so must the conjugate quantities. Now note that the
fifth coordinate $\lambda $ is qualitatively different from the other four $%
x^{^\mu }$; look for example at the metric (2.1). Further, the symmetry
group \textbf{C} includes translation groups on $\mathbf{r\ }$and $t$, hence
momentum and energy are conserved in particle scattering [11]. But there is
no translation group on $\lambda $ [7], so the conjugate quantity mass need
not be conserved.

The mass spectrum analyzed in \textbf{Sec. 3} does fit the experimental
numbers for the quarks, at least to within their (very loose) bounds.
However, this spectrum is not intended to be final and quantitative at this
stage. We only meant to show here that this particular 5-D theory required
by conformal symmetry is capable of predicting few-point mass spectra of the
right order of magnitude. The main problem is the crudity of the self-force
(3.1) adopted.\ This field does not in fact satisfy the boson field
equations (2.7) (see Ref. [7]) and must therefore be thought of as an
approximation to an actual solution$\footnote{%
The boson field equations (2.7) have the Coulombic solution $A_0=-g^{\prime
}\diagup \sqrt{\lambda ^2-r^2},$ $0\leq r<\lambda ;$ $=-g^{\prime }\diagup 
\sqrt{r^2-\lambda ^2},$ $0<\lambda <r<\infty $, other $A_{_\alpha }\equiv 0$.%
}$ or simply as a model. A quantitative theory needs a realistic self-force,
perhaps one involving also $SU(2)$ gauge bosons.

A few other points, including some puzzles, will be mentioned.

\noindent 1)\qquad The signature $\sigma =-$ was needed for an interesting
mass spectrum. We can show that for $\sigma =+$ a one-point spectrum results
for $\alpha >0$ (unpublished). The puzzle here is that $\sigma =+$ is
definitely indicated in the classical self-force theory [7], which
successfully resolves the anomalies due to classical point particles.

\medskip \noindent 2)\qquad Notice that if the lepton self-force is
electromagnetic: $\alpha \approx 1/137,$ the mass spectrum (3.4) cannot fit
the $T_3=-1/2$ leptons $e$, $\mu $, and $\tau $ since then $\gamma \equiv
(\alpha ^2-\nu ^2)^{1/2}$ is pure imaginary for $1<\nu ^2<2$. This is a
puzzle. But we add that for $\sigma =+,$ $(\nu ^2-\alpha ^2)^{1/2}$ occurs
where $(\alpha ^2-\nu ^2)^{1/2}$ occurs for $\sigma =-$, hence the equation
(3.2) written for $\sigma =+$ with a better self-force than (3.1) might work.

\noindent 3)\qquad For \textit{perfectly free} spin 1/2 fermions (no
external force \textit{and} no self-force) the field equation (2.4) with $%
A_{_\alpha }\equiv 0$, $\sigma =+$ or $-$, space-time dependence in $%
e^{ip\cdot x}$ with $p^2+m^2=0$, and $\gamma _{_\alpha }$ and $\gamma $ in
the EW representation is easily solved. The $\lambda $-dependence is in
factors $\lambda ^{5/2}Z_{\mu _L}(m\lambda )$ and $\lambda ^{5/2}Z_{\mu
_R}(m\lambda )$ for the $L$- and $R$-handed components of $\psi $, with $\mu
_L\neq \mu _R$. The $Z_{_\mu }$ are cylinder functions of order $\mu $. The
mass spectrum is continuous, $0\leq m<\infty $. In the case $\sigma =+$ if $%
\nu =-1/2$ is chosen for the Casimir invariant, then in the limit $%
m\rightarrow 0$ (neutrino solution) only a left-handed neutrino survives.
This makes the value $\nu =-1/2$ very attractive theoretically for leptons.
Perfectly free fermions are unphysical because of the continuous mass
spectrum. But this also supports the idea that the mass problem for leptons
should be phrased in the space $\sigma =+$ (\textit{cf.} point (2) above)
with $\nu =-1/2$.

\noindent 4)\qquad As indicated briefly above, this theory based on \textbf{C%
} instead of \textbf{P} as the kinematical symmetry group of particle
physics is compatible with the EW theory. Further, it furnishes a
theoretical foundation for some of the features of that theory adopted on
the basis of experiment. Consider the following points. (a) The six basic
anticommuting $\gamma $-matrices (2.3a) demand an 8-dimensional spinspace,
thus allowing the upper and lower 4-spinors to be identified with the $%
T_3=\pm 1/2$ isodoublets. (b) But more than this, in the differential
operator involving the primary gauge bosons B$_{_\alpha }$ and W$_{_\alpha
}^{}$ $^i$\newline
$(i=1,2,3)$, the spin algebra of the $SU(2)\times U(1)$ internal symmetry
group is formed entirely from the $8\times 8$ $\gamma $-matrices (2.3a,b).
Define the matrices 
\begin{equation}
\tau _i\equiv \gamma \text{ , }\tau _2\equiv i\gamma \beta _7\text{ , }\tau
_3\equiv \beta _7\text{ .}  \tag{4.1}
\end{equation}
Then these have the same commutation relations as the Pauli matrices.
Further, in the EW representation (2.5) they take exactly the standard form,
where the 1's and 0's are $4\times 4$. Contrast this with the situation in
the present day EW theory where generators of the internal symmetry group $%
SU(2)$, unrelated to the $\gamma _{_\mu }$, are imported from the outside.
The handedness projections $P_{h^{\prime }}$, $h^{\prime }=\pm $, are built
from the $8\times 8$ $H\equiv \lambda \beta _7\widetilde{\gamma }_5$, which
takes the form 
\begin{equation}
H=\left( 
\begin{array}{ll}
h & 0 \\ 
0 & h
\end{array}
\right) \text{ ,}  \tag{4.2}
\end{equation}
where $h$ is the $4\times 4$ handedness operator (see below Eq. (2.6)), in
the EW representation. (c) If the Lagrangian 
\begin{equation}
\mathcal{L}=\bar{\psi}\left[ \tilde{\gamma}\cdot D+\tilde{\gamma}\text{ }^5%
\text{ }D_5+2\sigma \lambda \tilde{\gamma}_5+\nu \beta _7\right] \psi \text{
,}  \tag{4.3}
\end{equation}
which yields the field equation (2.4), is equipped with the gauge bosons $%
B_{_\alpha }$ and $W_{_\alpha }$ $^i$, it exactly reproduces the Lagrangian
of the EW theory ([1], Chap. 7) plus some extra terms coming from the fifth
components $B_5$ and $W_5$ $^i$, presumably small corrections to the $4-D$
theory. Then the standard mixing produces the photon and $Z$ fields. d)
However, the aspect in which this theory is \textit{not} compatible with the
EW theory (or the whole Standard Model) is the main point of this paper. In
this theory the fermions may be massive, like the quarks considered in this
paper. The fifth dimension plus an appropriate self force provides the
masses. The Higgs mechanism is unnecessary.

\pagebreak

\noindent \textbf{APPENDIX. SOLUTION\ FOR\ THE\ MASS\ EIGENSTATES\ AND\
SPECTRUM}

\bigskip

Insert the formally $2\times 2$ representation 
\begin{equation}
\gamma ^0=i\left( 
\begin{array}{ll}
0 & 1 \\ 
1 & 0
\end{array}
\right) \text{, }\gamma _r=i\left( 
\begin{array}{ll}
0 & -1 \\ 
1 & 0
\end{array}
\right) \text{, }h=\left( 
\begin{array}{ll}
1 & 0 \\ 
0 & -1
\end{array}
\right)  \tag{A1}
\end{equation}
and $g(\lambda )=\left( 
\begin{array}{l}
F \\ 
G
\end{array}
\right) $ into Eq. (3.3). $F$ and $G$ are thus 2-spinors; in fact $F=g_R$
and $G=g_L$ in view of the form (A1) of the handedness operator $h$.
Multiplying by $-i$ we get 
\begin{equation}
\begin{array}{l}
(M-i\kappa -\alpha /\lambda )G-i(\partial _{_\lambda }+(i\nu -2)/\lambda )F=0%
\text{ ,} \\ 
\\ 
(M+i\kappa -\alpha /\lambda )F+i(\partial _{_\lambda }-(i\nu +2)/\lambda )G=0%
\text{ .}
\end{array}
\tag{A2}
\end{equation}
Rephase: $iF\rightarrow F$, $G\rightarrow G$. Define 
\begin{equation}
\beta _1\equiv M+i\kappa \text{, }\beta _2\equiv M-i\kappa \text{, }\beta
^2\equiv \beta _1\beta _2=M^2+\kappa ^2\text{.}  \tag{A3}
\end{equation}
Divide equations $(A2)$ by $\beta \equiv \sqrt{\beta ^2}$ and put $\beta
\lambda \equiv \tau $. 
\begin{equation}
\begin{array}{l}
(\beta _2/\beta -\alpha /\tau )G-(\partial _{_\tau }+(i\nu -2)/\tau )F=0%
\text{ ,} \\ 
\\ 
(\beta _1/\beta -\alpha /\tau )F-(\partial _{_\tau }-(i\nu +2)/\tau )G=0%
\text{ .}
\end{array}
\tag{A4}
\end{equation}
Set $F,G\equiv e^{-\tau }(f,g)$. Then $\partial _{_\tau }F=(\dot{f}%
-f)e^{-\tau }$ etc. where $^{\bullet }\equiv \partial /\partial \tau $.
Solve the equations in terms of $f$ and $g$ $\,$by the power series 
\begin{equation}
f=\tau ^s\sum\limits_{n=0}^\infty a_n\tau ^n\text{ , }g=\tau
^s\sum_{n=0}^\infty b_n\tau ^n\text{, }a_0\text{ and }b_0\neq 0\text{ .} 
\tag{A5}
\end{equation}
When these power series are inserted into the equations for $f$ and $g$ and
coefficients of $\tau ^{s+n-1}$ equated to $0$, we obtain 
\begin{equation}
\begin{array}{l}
(\beta _2/\beta )b_{n-1}-\alpha b_n-(s+n)a_n+a_{n-1}-(i\nu -2)a_n=0\text{ ,}
\\ 
\\ 
(\beta _1/\beta )a_{n-1}-\alpha a_n-(s+n)b_n+b_{n-1}+(i\nu +2)b_n=0\text{ .}
\end{array}
\tag{A6}
\end{equation}
Multiply the top equation (A6) by $\beta _1/\beta $ and subtract the bottom
equation. The terms $a_{n-1}$ and $b_{n-1}$ go out since $\beta _1\beta
_2/\beta ^2=1$. After rearrangement this gives 
\begin{equation}
\left[ (\beta _1/\beta )(s+n-2+i\nu )-\alpha \right] a_n=\left[ s+n-2-i\nu
-\beta _1\alpha /\beta _2\right] b_n\text{.}  \tag{A7}
\end{equation}
To get the indicial equation choose $n=0$ in Eq. (A6) and ignore the terms $%
a_{-1}$ and $b_{-1}$. The determinant must vanish so that nonzero $a_0$ and $%
b_0$ result; the result is 
\begin{equation}
S_{_\eta }\equiv s_{_\eta }-2=\eta (\alpha ^2-\nu ^2)^{1/2}\text{, }\qquad
\eta =\pm \text{ .}  \tag{A8}
\end{equation}
(We have changed the subscript $\tau $ on $S_{_\tau }$, Eq. (3.4b), to $\eta 
$ so as not to confuse it with the $\tau \equiv \beta \lambda $ of Eq. (A4) 
\textit{et seq.)} This is Eq. (3.4b). By letting $n\rightarrow \infty $ in
Eq. (A7) we get $b_n=(\beta _1/\beta )a_n$ in this limit; substituting this
into both equations (A6) for $n\rightarrow \infty $, we find $%
a_n/a_{n-1}=2/n $ and the same for the $b$'$s$ in this limit. Thus both
series (A5) diverge like $e^{2\tau }$, which is not allowed by the assumed
finiteness of the norm. Hence both series must terminate: 
\begin{equation}
a_{n^{\prime }+1}=b_{n^{\prime }+1}=0\text{ , }n^{\prime }=0,1,2,\cdots 
\text{ .}  \tag{A9}
\end{equation}
Set $n=n^{\prime }+1$ in Eq. (A6); we get $b_{n^{\prime }}=-(\beta _1/\beta
)a_{n^{\prime }}$ .\ Put this result into Eq. (A7) for $n=n^{\prime }$.
After cancellation of some terms and rearrangement 
\begin{equation}
2(\beta _1/\beta )(s+n^{\prime }-2)-\alpha (1+(\beta _1/\beta )^2)=0 
\tag{A10}
\end{equation}
results. Divide this by $2\beta _1/\beta $ and use $\beta _1/\beta =\beta
/\beta _2$. After some algebra we obtain 
\begin{equation}
S_{_\eta }+n^{\prime }=\alpha M/\beta \text{ .}  \tag{A11}
\end{equation}
(This implies Eq. (3.4c).) Finally, do some algebra on Eq. (A11), using $%
\beta \equiv \sqrt{M^2+\kappa ^2}$, to solve for $M$. This gives the mass
spectrum (3.4a).

\textit{The mass eigenstates.} From \textbf{Sec. 3} and this Appendix, the
mass eigenstates are $\psi =\left( 
\begin{array}{l}
F \\ 
G
\end{array}
\right) $, where the $2$-spinors $F$ and $G$ are 
\begin{equation}
F(t,r,\lambda )=e^{-iMt}\text{ }e^{-\kappa r}\text{ }F(\lambda )\text{ ,
\quad }G(t,r,\lambda )=e^{-iMt}\text{ }e^{-\kappa r}G(\lambda )\text{ ,} 
\tag{A12}
\end{equation}
\begin{equation}
\begin{array}{l}
F(\lambda )=(-i)e^{-\tau }\tau ^{s_{_\eta }}\sum\limits_{n=0}^{n^{\prime
}}a_n\tau ^n\times u_{+}\text{ ,} \\ 
\\ 
G(\lambda )=e^{-\tau }\tau ^{s_{_\eta }}\sum\limits_{n=0}^{n^{\prime
}}b_n\tau ^n\times u_{-}\text{ ,}
\end{array}
\tag{A13}
\end{equation}
where 
\begin{equation}
\tau \equiv \beta \lambda =\left[ (M_q/\kappa )^2+1\right] ^{1/2}\kappa
\lambda \text{ .}  \tag{A14}
\end{equation}
Here the quantum number of the eigenstate $q\equiv (n^{\prime },\eta )$ and $%
M_q/\kappa $ is given by Eq. (3.4a) with the sign $\tau $ changed to $\eta $%
. The relation of the $b_n$ to $a_n$ and the $a_n$ to the $a_{n-1}$ are
given by Eqs. (A7) and (A6). The constant $2$-spinors $u_{+}$ and $u_{-}$
are normalized in some way; the overall normalization of the $4$-spinor $%
\psi $ is secured by the free parameter $a_0$.

\bigskip

\noindent \textbf{REFERENCES}

\medskip

\noindent [1] G. Kane, \textit{Modern Elementary Particle Physics}
(Addison-Wesley, New York, 1993),\ Chap. 8.

\noindent [2] P.C.W. Davies and J.\ Brown, \textit{Superstrings, a Theory of
Everything?} (Cambridge University\ Press, Cambridge, 1988).

\noindent [3] B.\ Greene, \textit{The Elegant Universe} (Norton, New York,
1999).

\noindent [4] V. Bargmann and E.P.\ Wigner, Proc. Nat. Acad. Sci\textit{.} 
\textbf{34}, 211 (1948).

\noindent [5] F. Klein, Math.\ Ann\textit{.} \textbf{5}, 257 (1872).
\_\_\_\_\_\_\_, \textit{Vorlesungen \"{u}ber H\"{o}here Geometrie 3 Aufl},
(Springer, Berlin 1926).

\noindent [6] A.\ Einstein, \textit{The Meaning of Relativity}, 4th ed.
(Princeton University Press, Princeton 1953).

\noindent [7] R. L. Ingraham, Int. J. Mod. Phys. \textbf{7}, 603 (1998).

\noindent [8) R. L. Ingraham, Nuovo Cimento \textbf{68B}, 203, 218 (1982).

\noindent [9] L.I. Schiff, \textit{Quantum Mechanics, 2nd ed.} (McGraw-Hill,
New York 1955), Sec. 44.

\noindent [10] Review of Particle Physics, Euro. Phys. J. \textbf{C3}
(1998), p. 24.

\noindent [11] J.M. Jauch and F. Rohrlich, \textit{The Theory of Photons and
Electrons} (Addison-Wesley, Cambridge U.S.A. 1955), Sec. 1-11.

%\newpage

%\noindent \textbf{FIGURE\ CAPTIONS}

\begin{figure}[p]

\epsfysize=200mm
\vspace*{-1in}
\epsffile{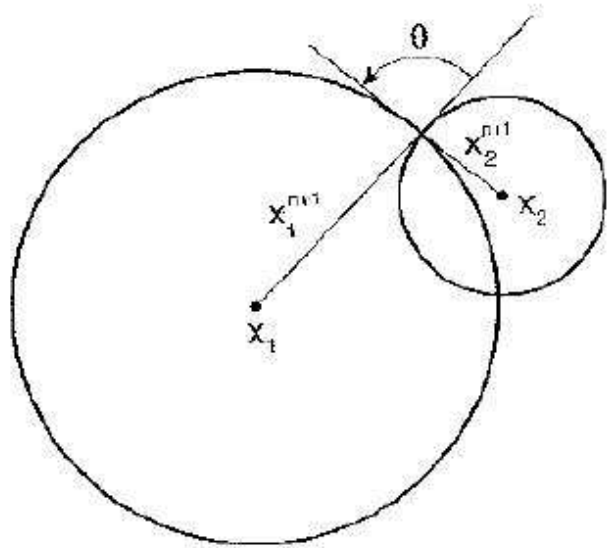}

\caption{The spheres $x_{_1}^{^{\;\alpha }}$ and $%
x_{_2}^{^{\;\alpha }},$ $\alpha =1,2,\cdots $ $n,$ $n+1$, in $E^{^n}$
intersecting under angle $\theta $. Here the center $\mathbf{x}_{_1}$ stands
for $\{x_{_1}^{\;1}$, $x_{_1}^{\;2}$, $\cdots $ $x_{_1}^{^{\;n}}\}$ and
similarly for $\mathbf{x}_{_2}$.}
\end{figure}
\end{document}